\documentclass[10pt,aps,showpacs,twocolumn]{revtex4}
\usepackage{amssymb}

\usepackage{epsfig}
\usepackage{amsmath}
\usepackage{bm}
\usepackage{epsfig}

\def\be{\begin{equation}}
\def\ee{\end{equation}}
\def\bea{\begin{eqnarray}}
\def\eea{\end{eqnarray}}

\begin{document}

\title{{\Large Connection between Wigner distribution of nonclassical fields and
collapse-revival of atomic dynamics}}
\author{C. H. Raymond Ooi}
\affiliation{\sl Department of Physics, Korea University,
Anam-dong, Seongbuk-gu, Seoul, 136-713 Republic of Korea}

\date{\today}

\begin{abstract}
The transient evolution of nonclassical radiation fields
interacting with an atom in a cavity is correlated to the atomic
dynamics. Connection between the atomic phases of collapse and
revival, and the Wigner function pattern is explored. Initial
classical coherent field can evolve into nonclassical field.
Schrodinger cat state field is generate during the atomic collapse
phase. For initial Schrodinger cat field state, the state
dissolves with time but revives during the atomic phase that
corresponds with the initial phase. More intricate but orderly
characteristics are found between the collapse-revival dynamics
and the patterns of the Wigner function for initial thermal field.
\end{abstract}

\maketitle

\textit{Introduction}

Schr\"{o}dinger's cat state $|\alpha \rangle \pm e^{i\phi }|-\alpha \rangle $%
, a superposition of classical (coherent) states, is a nonclassical state.
One of the challenges in quantum communication science is to produce the
state with large $|\alpha |^{2}$. Six atoms atomic cat state \cite{six atoms
Sch cat} has also been produced. The cat state for fields may be produced
from squeezed sources, linear processes, simple photon counting scheme \cite
{Ralph cat}, and also by conditional measurement \cite{Cat by cond meas}.
Single photon subtraction from a squeezed vacuum state can generate
Schr\"{o}dinger's kitten (small $|\alpha |^{2}$) via a technique which
involves homodyne detection \cite{Grangier Sc}. Recently, large-amplitude
coherent-state superposition has been generated via photon subtraction \cite
{large coh}. These are promising developments toward macroscopic quantum
superposition of states for quantum information processing.

Photon substraction via conditional measurement can improve the quality of
teleportation with continuous variables \cite{teleport with photon sub}.
Photon subtraction of squeezed vacuum gives coherent superposition of odd
photon number states with nonclassicality\cite{Polzik}, such as negativity
of the Wigner function and sub-Poissonian statistics \cite{Biswas Agarwal}.
Entanglement between two pulses that are Gaussian quadrature-entangled can
also be increased by coherent subtraction of single photons \cite{Grangier
increase ent}.

Conversely, $m$-photon addition of coherent state produces phase squeezing,
sub-Poissonian statistics and negativity of the Wigner function \cite
{Agarwal add coh}. Similar nonclassical character is found in photon-added
even and odd coherent states (positive and negative Schr\"{o}dinger's cat)
\cite{Dodonov +- cat}. For time dependent electromagnetic field frequency,
photon addition of coherent state gives enhanced dynamical squeezing \cite
{Dodonov photon add coh}. In general, both the subtraction and addition of
single photon give nonclassical results.

Since photon addition (subtraction) correspond to downward (upward)
transition in atomic dynamics, the atom and the field are closely
correlated. Thus, the distinct mechanisms between photon addition and
subtraction can be better understood by looking at the atomic dynamics of a
single atom-single mode cavity field system. Such cavity system has
displayed quantum effects like Rabi oscillations, collapse-revival dynamics
\cite{collapse revival Meystre} and atom-field entangled states \cite
{Haroche}.

In this letter, we explore the nature of nonclassical fields by studying the
atomic dynamics, particularly the connection between the nonclassical
behavior of the Wigner function and the phenomena of collapse-revival or
atomic inversion. Although such system was studied by ref. \cite{photon
added thermal state}, it was confined only photon addition process on single
mode thermal field and weak coupling (short time) regime $gt<<1$. Our
present work is valid for arbitrary time. One interesting result is that the
Schr\"{o}dinger cat state can be generated from coherent state during the
collapse in the atomic inversion. The results in this work provide insights
into the subtle effect of atomic quantum interference on nonclassical light,
photon subtraction and addition processes; and lead to the possibility of
controlling the quantum state of the field by controlling the atomic
dynamics.

\textit{Atom-field dynamics in cavity}

The single mode-single atom (two levels $a$ and $b$) dynamics is governed by
the interaction Hamiltonian $V=\hbar g(\sigma _{+}ae^{i\Delta t}+\sigma {-}%
a^{+}e^{i\Delta t})$ (in interaction picture) with $\sigma _{+}=|a\rangle
\langle b|$, and $\sigma _{-}=|b\rangle \langle a|$ and $~\Delta =\omega
-\nu $. Consider that initially the atomic state $\rho ^{a}(0)$ is
uncorrelated to the field $\rho (0)$, so $\hat{\rho}(0)=$ $\hat{\rho}%
_{f}(0)\otimes \hat{\rho}_{a}(0)$. The evolution of the atom-field system
evolves according to $|\psi (t)\rangle =\sum_{n}[C_{a,n}(t)|a,n\rangle
+C_{b,n}(t)|b,n\rangle ]$ with the coefficients given by\cite{QO} $%
C_{a,n}(t)=e^{i\Delta
t/2}[C_{a,n}(0)r_{n}-iC_{b,n+1}(0)q_{n}],C_{b,n}(t)=e^{-i\Delta
t/2}[C_{b,n}(0)r_{n-1}^{\ast }-iC_{a,n-1}(0)q_{n-1}]$ where $r_{n}=\cos (%
\frac{\Omega _{n}t}{2})-i\frac{\Delta }{\Omega _{n}}\sin (\frac{\Omega _{n}t%
}{2})$, $q_{n}=\frac{2g\sqrt{n+1}}{\Omega _{n}}\sin (\frac{\Omega _{n}t}{2})$
and $\Omega _{n}^{2}=\Delta ^{2}+4g^{2}(n+1)$. The initial coefficients are $%
C_{a,n}(0)=C_{a}(0)C_{n}(0)$, $C_{b,n+1}(0)=C_{b}(0)C_{n+1}(0)$ satisfy $%
|C_{a}(0)|^{2}+|C_{b}(0)|^{2}=1$.

In general, the matrix elements of the field is obtained by tracing out the
atomic system (a), $\rho _{nm}=\langle n|\hat{\rho}_{f}|m\rangle
=\sum\limits_{s}\langle n|\hat{\rho}_{ss}|m\rangle $ where $\hat{\rho}%
_{f}=Tr_{a}\{\hat{\rho}(t)\}=\sum\limits_{s=a,b}\hat{\rho}_{ss}$ $\hat{\rho}%
_{ss}=\langle s|\hat{\rho}|s\rangle ,\langle n|\hat{\rho}_{ss}|m\rangle
=C_{s,n}(t)C_{s,m}^{\ast }(t)$ ($s=a,b$).

\textit{Field density matrix elements}

From the full expressions for the coefficients $C_{a,n}(t)$ and $C_{b,n}(t)$%
, the transient density matrix elements of the field can be obtained from
\begin{eqnarray}
\rho _{nm}(t) &=&\langle n|\{\hat{\rho}_{aa}(t)+\hat{\rho}_{bb}(t)\}|m\rangle
\notag \\
&=&C_{a,n}(t)C_{a,m}^{\ast }(t)+C_{b,n}(t)C_{b,m}^{\ast }(t)
\end{eqnarray}
Since initial atomic state is not correlated to the field, $%
C_{x,n}(0)C_{y,n^{\prime }}^{\ast }(0)=C_{x}(0)C_{y}^{\ast }(0)\rho
_{nn^{\prime }}(0)$ we may write the expression for $\rho _{nm}(t)$ that is
general valid, even for mixed field states

\begin{widetext}
\begin{eqnarray}
\rho _{nm}(t) &=&\{|C_{a}(0)|^{2}r_{n}r_{m}^{\ast
}+|C_{b}(0)|^{2}r_{m-1}r_{n-1}^{\ast }\}\rho
_{nm}(0)+|C_{b}(0)|^{2}q_{n}q_{m}\rho
_{n+1,m+1}(0)+|C_{a}(0)|^{2}q_{n-1}q_{m-1}\rho _{n-1,m-1}(0)+  \notag \\
&&iC_{a}(0)C_{b}^{\ast }(0)\{r_{n}q_{m}\rho
_{n,m+1}(0)-q_{n-1}r_{m-1}\rho _{n-1,m}(0)\}+iC_{b}(0)C_{a}^{\ast
}(0)\{r_{n-1}^{\ast }q_{m-1}\rho _{n,m-1}(0)-q_{n}r_{m}^{\ast
}\rho _{n+1,m}(0)\}  \label{pnm general}
\end{eqnarray}
\end{widetext}

where $\rho _{nm}(0)=\langle n|\hat{\rho}_{f}(0)|m\rangle $ with $\hat{\rho}%
_{f}(0)$ being the initial state of the field. This expression will be used
to compute the Wigner function for the field.

\textit{Atomic inversion }

The dynamics of atomic inversion $n_{ab}=\sum_{m}^{\infty
}\{|C_{a,m}(t)|^{2}-|C_{b,m}(t)|^{2}\}$ is computed by tracing over the
photon number states, using
\begin{widetext}
\begin{equation}
n_{ab}=\sum_{m}^{\infty }\left[
\{|C_{a}(0)r_{m}(t)|^{2}-|C_{b}(0)r_{m-1}(t)|^{2}%
\}p_{m}(0)+|C_{b,m+1}(0)q_{m}|^{2}p_{m+1}(0)-|C_{a}(0)q_{m-1}|^{2}p_{m-1}(0)%
\right]
\end{equation}
\end{widetext}

\textit{Wigner function versus density matrix elements}

Nonclassical states of light have been studied through various physical
parameters; the most typical ones being squeezing, antibunching in $G^{(2)}$%
, entanglement criteria, Mandel's Q and Wigner's function. The squeezing and
sub-Poissonian may not serve reliably to quantifying nonclassicality\cite
{Lee}, \cite{Tara}. Negativity of the Wigner function is a more reliable
quantity and has been used to study the concept of photon addition and
subtraction processes for producing nonclassical states. The relation
between the Wigner function and the density matrix elements for the field $%
\rho _{nm}$ can be derived\cite{QO} by using the identities found in \cite
{Cahill Glauber}, $\ $
\begin{eqnarray}
W(\alpha ,\alpha ^{\ast },t) &=&\frac{2e^{2\left| \alpha \right| ^{2}}}{\pi
^{2}}\sum_{m,n}^{\infty }\rho _{nm}(t)\times  \label{W relate pnm} \\
&&\int \langle -\beta |n\rangle \langle m|\beta \rangle e^{2(\beta ^{\ast
}\alpha -\beta \alpha ^{\ast })}d^{2}\beta  \notag
\end{eqnarray}
where $\hat{\rho}_{f}(t)=\sum_{m,n}^{\infty }|n\rangle \rho _{nm}(t)\langle
m|$ for the field in photon number basis. After some calculations using $%
\langle n|\alpha \rangle =e^{-|\alpha |^{2}/2}\frac{\alpha ^{n}}{\sqrt{n!}}$
and $\int \int \beta ^{m}\beta ^{\ast n}e^{-|\beta |^{2}}e^{(\beta ^{\ast
}2\alpha -\beta 2\alpha ^{\ast })}d^{2}\beta =\pi L_{n}^{m-n}(|2\alpha
|^{2})n!e^{-|2\alpha |^{2}}(2\alpha )^{m-n}$ we finally obtain

\begin{equation*}
W(\alpha ,\alpha ^{\ast },t)=\frac{2e^{-2\left| \alpha \right| ^{2}}}{\pi }%
[\sum_{m=0}^{\infty }(-1)^{m}L_{m}^{0}(|2\alpha |^{2})\rho _{mm}(t)+
\end{equation*}
\begin{equation}
\sum_{m=1}^{\infty }\sum_{n=0}^{m-1}(-1)^{n}\sqrt{\frac{n!}{m!}}%
L_{n}^{k}(|2\alpha |^{2})\{(2\alpha )^{k}\rho _{nm}(t)+\text{c.c.}\}]
\label{W final}
\end{equation}
where $k=m-n>0$, $L_{n}^{k}(x)$ are the associated Laguerre polynomials and
the first term adds up the diagonal elements. This is the main equation, to
be used along with Eq. \ref{pnm general} for obtaining the results below.

\textit{Initial coherent state}

For coherent state, $|\alpha _{0}\rangle =\sum\limits_{n}f_{n}|n\rangle \ $%
with $f_{n}=e^{-\left| \alpha _{0}\right| ^{2}/2}\frac{\alpha _{0}^{n}}{%
\sqrt{n!}}$. For initial Schr\"{o}dinger's cat states$\ |\psi \rangle
_{cat}=N(|\alpha _{0}\rangle \pm e^{i\phi }|-\alpha _{0}\rangle
)=N\sum\limits_{n}f_{n}|n\rangle $ with $N=\frac{1}{\sqrt{2(1+e^{-2|\alpha
_{0}|^{2}}\cos \phi )}}$, we have $f_{n}=$ $e^{-\left| \alpha _{0}\right|
^{2}/2}\frac{\alpha _{0}^{n}\pm (-\alpha _{0})^{n}}{\sqrt{n!}}$. The states
are also referred to as even/odd superposition of coherent states. Thus, in
general $\hat{\rho}_{f}=|\alpha _{0}\rangle \langle \alpha _{0}|=e^{-|\alpha
_{0}|^{2}}\sum\limits_{n}\frac{\alpha _{0}^{n}\alpha _{0}^{\ast m}}{\sqrt{%
n!m!}}|n\rangle \langle m|=\sum\limits_{n}\rho _{nm}(0)|n\rangle \langle m|$
gives

\begin{equation}
\rho _{nm}(0)=f_{n}f_{m}^{\ast }(0)  \label{pnm(0) coh state}
\end{equation}
where $\rho _{nm}(0)=e^{-\left| \alpha _{0}\right| ^{2}}\frac{\alpha
_{0}^{n}\alpha _{0}^{\ast m}}{\sqrt{n!m!}}$ and $e^{-\left| \alpha
_{0}\right| ^{2}}\frac{(\alpha _{0}^{n}\pm (-\alpha _{0})^{n})(\alpha
_{0}^{\ast m}\pm (-\alpha _{0}^{\ast })^{m})}{\sqrt{n!m!}}$ for coherent
state and the Schr\"{o}dinger's cat states, respectively. Here, $\alpha
_{0}=re^{i\theta }$ and $r$ determines the distance between the centroid of
the Wigner function and the origin $\alpha =0$ while $\theta $ determines
the angle with respect to Re$\alpha $.

\begin{figure}[tbp]
\center\epsfxsize=8cm\epsffile{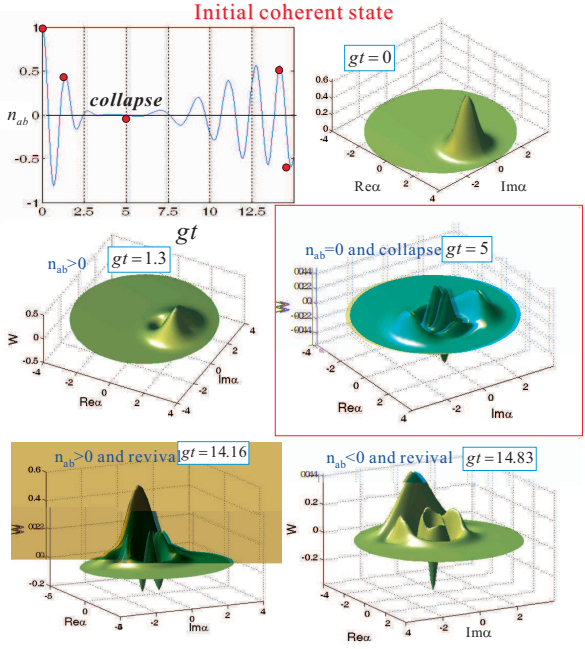}
\caption{(Color online) Initial coherent state. The tagged circles in the
first plot correspond to the times for the plots of the Wigner function $W$.
Parameters are $C_{a}(0)=1,\protect\alpha _{0}=1$.}
\label{coherent}
\end{figure}
\textit{Fields with diagonal photon number}

For fields with diagonal number states,

\begin{eqnarray}
\rho _{nm}(t) &=&\delta
_{mn}\{|C_{a}(0)r_{m}|^{2}+|C_{b}(0)r_{m-1}|^{2}\}p_{m}(0)  \notag \\
&&+|C_{a}(0)q_{m-1}|^{2}p_{m-1}(0)+|C_{b}(0)q_{m}|^{2}p_{m+1}(0)\}  \notag \\
&&+\delta _{n,m-1}iC_{b}(0)C_{a}^{\ast }(0)\{r_{n-1}^{\ast }q_{m-1}p_{m-1}(0)
\notag \\
&&-r_{m}^{\ast }q_{n}p_{m}(0)\} \label{pnm diagonal}
\end{eqnarray}

where we have abbreviated the diagonal matrix elements of the field by $%
p_{m-1}(0)=\rho _{m-1,m-1}(0)$.

Note that at $t=0$ we have $W(\alpha ,\alpha ^{\ast },0)=\frac{2e^{-2\left|
\alpha \right| ^{2}}}{\pi }\sum_{n}^{\infty }(-1)^{n}L_{n}^{0}(|2\alpha
|^{2})p_{n}(0)$. However, for finite $t$ the emission and absorption
processes develop coherences between photon numbers, which are accounted by
the second line when $m=n+1$.

When $C_{a}(0)=1$ only the diagonal elements contribute $\rho
_{nm}(t)=\delta _{mn}\{|r_{n}(t)|^{2}p_{n}(0)+q_{n-1}(t)^{2}p_{n-1}(0)\}$
which gives
\begin{eqnarray}
W(\alpha ,\alpha ^{\ast },t) &=&\frac{2e^{-2\left| \alpha \right| ^{2}}}{\pi
}\sum_{n}^{\infty }(-1)^{n}L_{n}^{0}(|2\alpha |^{2})\times  \notag \\
&&[|r_{n}(t)|^{2}p_{n}(0)+q_{n-1}(t)^{2}p_{n-1}(0)]
\end{eqnarray}

For thermal state $p_{n}(0)\doteq (1-e^{-\beta \hbar v})e^{-n\beta \hbar v}=%
\frac{1}{(1+\bar{n})}(\frac{\bar{n}}{(1+\bar{n})})^{n}$ where $\overline{n}=%
\frac{1}{exp(\hbar \omega /k_{B}T)-1}$. When $\overline{n}$ is large, $%
q_{n-1}\simeq q_{n}$ and $p_{n}(0)\simeq p_{n-1}(0)$ are good
approximations, and since $r_{n}(t)=\cos (\frac{\Omega _{n}t}{2})$, $%
q_{n}(t)=\sin (\frac{\Omega _{n}t}{2})$ for $\Delta =0$, the Wigner function
becomes essentially time independent $W\simeq \frac{2e^{-2\left| \alpha
\right| ^{2}}}{\pi (1+\bar{n})}\sum_{m}^{\infty }(-\frac{\bar{n}}{(1+\bar{n})%
})^{m}L_{m}^{0}(|2\alpha |^{2})$.

\begin{figure*}[tbp]
\center\epsfxsize=13cm\epsffile{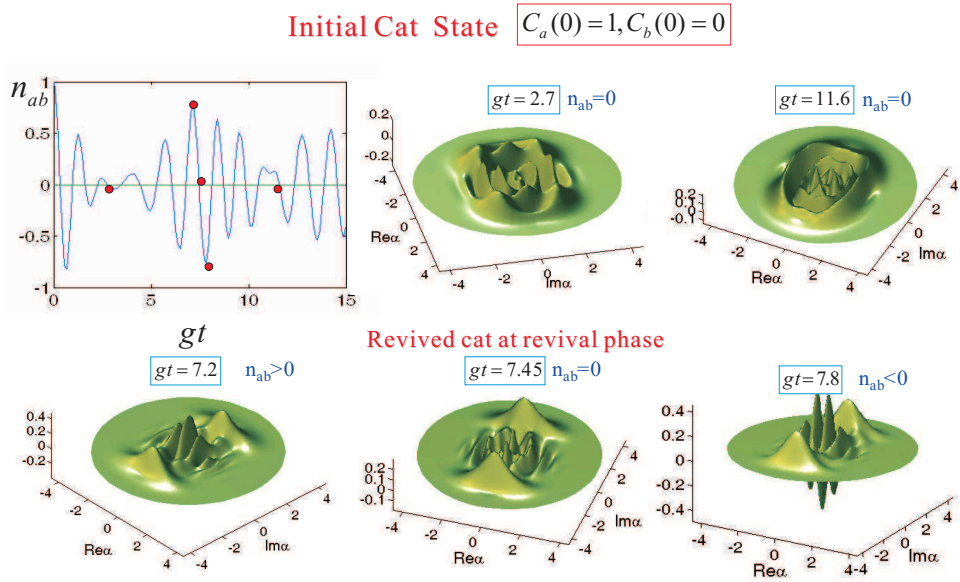}
\caption{(Color online) Initial field in Schrodinger cat state while atom is
excited. Parameters are $C_{a}(0)=1,C_{b}(0)=0,\protect\alpha _{0}=\protect%
\sqrt{5}$, $\protect\phi =0.$}
\label{Schcat}
\end{figure*}

\begin{figure*}[tbp]
\center\epsfxsize=17cm\epsffile{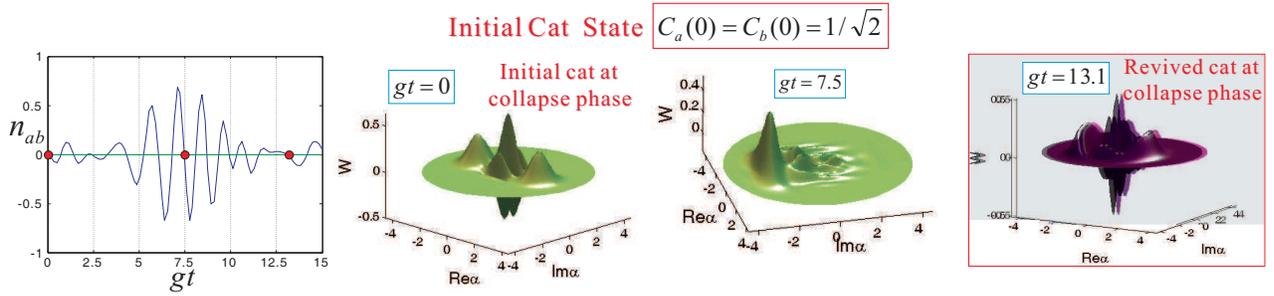}
\caption{(Color online) Initial field in Schrodinger cat state while atom is
in superposition of excited and ground states. Parameters are $%
C_{a}(0)=C_{b}(0)=1/\protect\sqrt{2},\protect\alpha _{0}=\protect\sqrt{5}$, $%
\protect\phi =0.$}
\label{Schcat2}
\end{figure*}
\textit{Results and Discussions}

We find some interesting properties by comparing the dynamics of the
inversion $n_{ab}(t)$ with the pattern in the Wigner function. For initial
coherent state, Fig. \ref{coherent} shows that the field evolves into the
Schr\"{o}dinger cat state $|\Psi \rangle _{cat}=N(|\alpha _{0}\rangle \pm
e^{i\phi }|-\alpha _{0}\rangle )$ when the atomic inversion collapse. At
points where $n_{ab}=0$, the Wigner patterns show the features that only
resemble the Wigner pattern for the cat state, i.e. two main peaks across $%
\alpha =0$ with interference fringes in between. However, in the region of
atomic inversion collapse, i.e. $n_{ab}=0$ for a finite period of time ($%
gt=2.6-6.2$ in \ Fig. \ref{coherent}), we find that the Wigner pattern is
\emph{identical} to the ideal Wigner pattern of the Schr\"{o}dinger cat.
This suggest that the atomic collapse phase is associated with photon
entanglement.

Let use now consider the initial Schr\"{o}dinger's cat state $|\Psi \rangle
_{cat}$. Figure \ref{Schcat} shows that the two main peaks of the cat state
''dissolve'' initially with time as the inversion begins to collapse. The
main peaks reappear as the result of rephasing, when the inversion revives,
as shown for $gt=7.2$. Even though $n_{ab}=0$ at $gt=7.45$, the Wigner
pattern shows the cat state since the atom is in the revival state. This
situation is reverse from the case with initial coherent state in Fig. \ref
{coherent}. Thus, the cat state is not always associated with the collapse
phase. Also, in general, the overall shape of the Wigner pattern depends on
the envelope of the inversion while the detailed signs of the peaks in the
oscillations depend on the sign of the inversion peaks.

We now consider the atom to be initially in the superposition of internal
states, $C_{a}(0)=C_{b}(0)=\frac{1}{\sqrt{2}}$ while the field is still in
the initial cat state. As shown in Fig. \ref{Schcat2}, the cat state is
found only during the collapse phase, which is the same as the initial
phase. Base on these observations, it may be conjectured that, the state of
the field returns to or close to the initial state, when the phase returns
to the initial phase at $t=0$. In other words, the revived field state
corresponds to the initial atomic phase.

\begin{figure*}[tbp]
\center\epsfxsize=17cm\epsffile{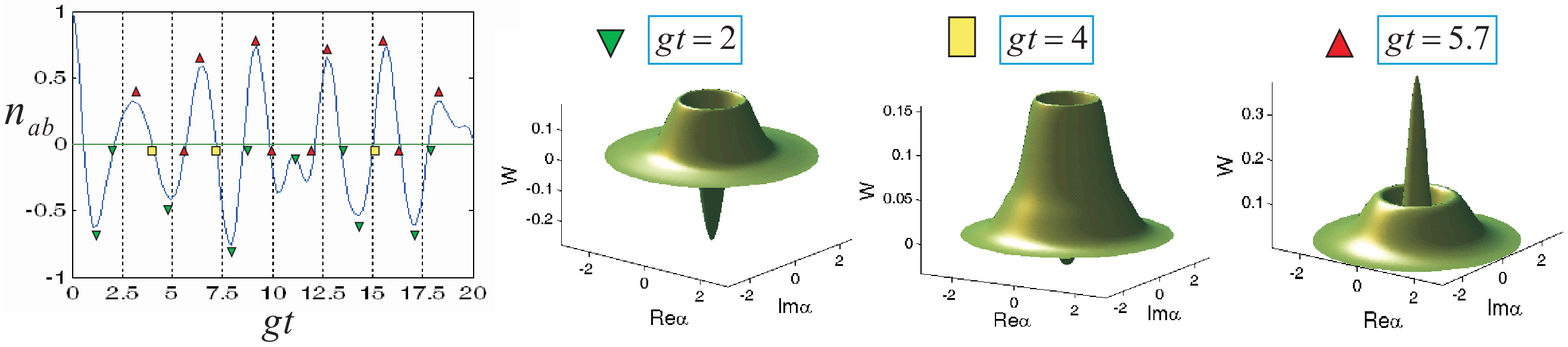}
\caption{(Color online) Initial thermal state. Parameters are $C_{a}(0)=1,%
\bar{n}=1$.}
\label{thermal}
\end{figure*}

In the case of thermal field (Fig. \ref{thermal}), the inversion is
connected to the Wigner function in a more sophisticated way and yet in an
orderly manner. Although there seems to be no region of collapse, the points
in $n_{ab}$ can be categorized three groups. First, at points with $n_{ab}=0$%
, the Wigner function shows an alternation between three distinct Wigner
patterns which we designate as $\nabla $, $\square $ and $\Delta $. Second,
at the positive peaks of $n_{ab}$, the Wigner pattern has only one typical
form, the $\Delta $ which looks like the ''Borobodur''. Third, the negative
peaks of $n_{ab}$ correspond to another typical Wigner pattern, the $\nabla $%
.

From the results we may conjecture two points in summary. First, an
entangled field state can be controlled by controlling the atomic dynamics.
Second, the atomic dynamics contain information about the nature the fields.
Verification of these results may involve application of tomography
technique \cite{Lvovsky} to map out the transient evolution of the Wigner
function in connection with the collapse-revival dynamics.


\begin{thebibliography}{99}
\bibitem{Lee}  C.T. Lee, \pra {\bf 44}, R2775 (1991).

\bibitem{Ralph cat}  T. C. Ralph \textit{et. al.}, \pra {\bf 68}, 042319
(2003).

\bibitem{Cat by cond meas}  M. Dakna \textit{et. al.}, \pra {\bf 55}, 3184
(1997).

\bibitem{Grangier Sc}  A. Ourjoumtsev \textit{et. al.}, Science 312, 83
(2006).

\bibitem{six atoms Sch cat}  D. Leibfried et. al., Nature 438, 639 (2005).

\bibitem{large coh}  Hiroki Takahashi et. al., \pra {\bf 101}, 233605 (2008).

\bibitem{teleport with photon sub}  T. Opatrny, G. Kurizki, and D.-G.
Welsch, \pra {\bf 61}, 032302 (2000).

\bibitem{Polzik}  J. S. Neergaard-Nielsen \textit{et. al.}, \prl {\bf 97},
083604 (2006).

\bibitem{Biswas Agarwal}  A. Biswas and G.S. Agarwal, \pra {\bf 75}, 032104
(2007).

\bibitem{Grangier increase ent}  A. Ourjoumtsev, R. Tualle-Brouri, and P.
Grangier, \prl {\bf 96}, 213601 (2006).

\bibitem{Agarwal add coh}  G. S. Agarwal, and K. Tara, \pra {\bf 43}, 492
(1991).

\bibitem{Dodonov +- cat}  V. V. Dodonov \textit{et. al.}, Quant. Semiclass.
Opt. 8, 413 (1996).

\bibitem{Dodonov photon add coh}  V. V. Dodonov \textit{et. al.},
\pra {\bf
58}, 4087 (1998).

\bibitem{collapse revival Meystre}  R. Kanamoto, E. M. Wright, and P.
Meystre, \pra {\bf 75}, 063623 (2007).

\bibitem{Haroche}  A Meunier \textit{et. al.}, \pra {\bf 74}, 033802 (2006).

\bibitem{photon added thermal state}  G. N. Jones, J. Haight and C. T. Lee,
Quant. Semiclass. Opt. 9, 411 (1996).

\bibitem{Tara}  G.S. Agarwal and K. Tara, \pra {\bf 46}, 485 (1992).

\bibitem{QO}  M. O. Scully and M. S. Zubairy, \textsl{Quantum Optics}
(Cambridge University Press, Cambridge, 1997).

\bibitem{Cahill Glauber}  K. E. Cahill and R. J. Glauber, Phys. Rev. 177,
1882 (1969); \textsl{Quantum theory of optical coherence: selected papers
and lectures}, R. J. Glauber (Wiley-VCH, Weinheim, 2007).

\bibitem{Lvovsky}  A. I. Lvovsky \textit{et. al.}, \prl {\bf 87}, 050402
(2001).
\end{thebibliography}
\end{document}